
\magnification 1200
\centerline {Relaxation to Spin Glass Ground State in One to Six Dimensions}
Dietrich Stauffer$^*$ and Paulo Murilo Castro de Oliveira

Institute of Physics, Fluminense Federal University\par
Av. Litor\^anea s/n, Boa Viagem, Niter\'oi, RJ 24210-340, Brazil

Abstract: In the $\pm J$ Edwards-Anderson spin glass, we find by Monte
Carlo simulation the (approximate) ground state energy. Also we check how
in the relaxation towards this ground state the fraction of never
flipped spins diminishes with time.

Spin glass simulations are now a classical problem of computational statistical
physics [1]. In the $\pm J$ Edwards-Anderson spin glass with energy
$\sum J_{ik} S_i S_k$, the nearest neighbor couplings $J_{ik}$ between the
spins $S_i = \pm 1$ are randomly $+J$ and $-J$ with equal
probabilities. The one-dimensional chain can be mathematicaly
transformed into a chain of Ising magnets, whereas in two and more dimensions
frustration appears. In three and more dimensions a phase transition
at finite temperature exists, which above six dimensions should
belong to the universality class of mean field theory. The present work
concerns the zero-temperature behavior only, which is also far from
trivial. A ground state here is meant to be a configuration which
does not have a measurably higher energy than the lowest possible energy.

Of the numerous ways to find the ground state (transfer matrix, simulated
annealing, ...), three were recently motivated by biological
analogies in the evolution of species: According to Darwinistic selection
of the fittest, random mutations lead to fitter individua. The maximum
fitness then corresponds to the minimum energy of the spin glass [2].
In all three approaches, a biological mutation corresponds to a spin flip.
Rodrigues and de Oliveira [3] combined energy-lowering
mutations with an oscillating
infinitesimal field to flip the ``free'' spins in its direction, i.e.
those spins whose energy
would not change by a flip in zero magnetic field; we refer to that paper
for a justification of these oscillations which are avoided here. Stauffer and
Stevens et al [4] allowed for rare mutations which increase the energy;
this method corresponds to simulated annealing in the limit of infinitesimally
low temperature. Sutton et al [5] combined these mutations with a genetic
algorithm, thus including sex which Stauffer [4] missed. In two dimensions,
ref.5 found within the error bars the desired ground state energy
whereas ref.4 without sex found only a slightly higher, metastable energy.
Thus ref.5 gave the best ground state energies.
The methods of ref.4 and 5 require a computer time growing with increasing
system size faster than the number of spins involved; therefore they and also
ref.3 used relatively small lattices. The aim of the present work is to
simplify the method of ref.3, to apply it to larger systems and higher
dimensions, and to see how the relaxation of these spin glasses compares
with that of Ising ferromagnets at zero temperature [6].

Suppose we have $n$ free spins in some configuration $S_0$, after all spins
on the square lattice with 3 or 4 surrounding excited (= broken) bonds
are flipped. This state is not a truely metastable state in the usual
sense: not necessarily all its neighboring states (differing from it by
only one flipped spin) have a higher energy. Instead, we have now $n$ such
neighboring states $S_1, S_2, \dots, S_n$ with the same energy as $S_0$.
Each one is reachable from $S_0$ by flipping just one of its $n$ free spins.
Any of these $n$ new states, called  $S_i$, can have some spin(s) surrounded
by three excited bonds, unlike $S_0$; this spin will be in the neighborhood
of the spin
which was flipped compared to $S_0$. If this is the case, the energy can
be further decreased simply by flipping this spin. Otherwise,
let $k$ be the number of further free spins of configuration $S_i$.
Now we have $k$ neighboring states
$S_{i,1}, S_{i,2}, \dots, S_{i,k}$ of $S_i$ with the same energy as $S_i$
and $S_0$. All of them are second neighbors of $S_0$ in configuration
space. Again, one of them may have some spins with three broken bonds,
allowing a further reduction of the energy, and so on. This means we have a
ramified set of possibilities for further energy minimization, always
through single mutations (one spin flip at each step) and never increasing
the energy (Darwinistic selection). The flips of free spins correspond to
mutations which do not change the fitness if taken alone, but which may
lead to increased fitness in combination with other, later mutations.
There are many ways to explore these bifurcating minimization paths along the
rugged energy landscape. Ref.3 adopted the iteratively reversed infinitesimal
magnetic field, and now we use the following simplified version.

Ref.3 flipped spins with a probability of 50 percent if and only if this flip
lowers the energy. To deal with those spins which otherwise would have a
vanishing flip energy (free spins), a homogeneous magnetic field was applied in
one direction until no more spin flips occured; then the field was reversed
until again equilibrium was reached, and this changing of the field direction
was repeated again and again until the resulting energy no longer changed
measurably. On the square lattice an energy of 0.16 was achieved, compared with
the literature value[1] of 0.15 and as good as the approximation 0.16 of
Stauffer[4]. (We normalize the energy to the interval between zero and unity,
with zero corresponding to the case where all bonds are satisfied, and unity to
the opposite case of all bonds broken.) Now we found it not necessary to waste
50 percent of the flips, and instead of flipping all free spins in the same
direction (of the homogeneous infinitesimal field) we flip them randomly, half
up and half down. Spins were updated sequentially by going through the lattice
like a typewriter. This algorithm thus corresponds to the zero-temperature
limit of Glauber or heat bath kinetics used also in ref.6: If a flip lowers the
energy it is always made, if it increases the energy it is never made, and if
it keeps the energy constant it is made with probability 1/2. As discussed in
ref.3, the flipping of the free spins gets the system out of many if not all
local energy minima which are obtained if only flips lowering the energy are
allowed; the latter method gives an energy of 0.24 instead of 0.15 on the
square lattice.

(By avoiding the method of oscillating homogeneous fields we
decreased only slightly the computer time per Monte Carlo step. But instead
of making 8000 iterations for one field direction, followed by
another 8000 for the opposite direction, and repeating this cycle
again and again, we now just use the Glauber or heat bath algorithm
at $T = 0$ for one stretch of 1000 (or more) iterations and extrapolated
from these data to infinite time. Thus lots of computer time was saved
compared with ref.3.)

Using one Fortran program working in one to six dimensions, we found the
one-dimensional energy to relax as $1/\sqrt t$ towards
$\pm 10^{-3}$, i.e. eventually all bonds are satisfied since no frustration
occurs. In higher dimensions the effective exponent for the time $t$ was
somewhat higher, and the energies extrapolated after $t = 1000$
sweeps through a lattice of a million spins were 0.158, 0.211, 0.290,
0.271, and 0.246 in two to six dimensions on hypercubic lattices.
In two and three dimensions we would have liked to have 0.150 and 0.20$_5$,
while we are not aware of previous ground state estimates in higher
dimensions. For smaller lattices, we note that slight further energy decrements
occur simply by taking larger and larger numbers of lattice sweeps.
Nevertheless this is prohibitively time consuming, and presumably still will
not lead to the true ground states.

(We made one run each for $L$ = 5 10$^6$, 2000, 150, 31, 17, and 10 in
one to six dimensions, 100 runs for $63^3, \, 15^4, \, 11^5$ and $7^6$,
and 4000 runs for $64^2$ and estimate the errors in our energies to
be of the order 0.001 or better.)

The fraction of spins which were never flipped decayed with time $t$ as
$t^{-0.37}$ towards zero in one dimension whereas in two to six dimensions this
fraction approached about one quarter, if we start with all spins up or all
spins random. This one-dimensional power law agrees with that of ferromagnets
[6], but in two and three dimensions the ferromagnets had this fraction also
going to zero.  Thus only for higher dimensions do spin glass and ferromagnet
agree again with a finite fraction of stable spins. Similarly to the two
dimensional case [3], we find the rms fluctuations of the energy varying
roughly as $L^{-d/2}$ also in higher dimensions, similar to random processes,
where $L$ is the system linear dimension.

The finite fraction of never flipped spins on the square lattice does not
seem to be due to insufficient computing time. For $L = 31$ we looked at
the number of never flipped spins after 10, 31, 100, 316, 1000, 3162,
10000, 31624, and 100000 sweeps through the lattice, and found it to
converge: 357, 287, 253, 212, 206, 204, 202, and 202, respectively,
amounting to between one fourth and one fifth of the 961 spins involved
here. Thus there is no constant decrement on this logarithmic time scale.

We believe that even for infinite times our method would not lead to the
true ground state. With our method the system leaves a metastable energy
minimum only if another configuration exists which can be reached by one spin
flip and which does not increase the energy. If we are in a minimum which
requires flipping at least one spin against the wishes of its neighbors,
then these neighbors prevent the system to leave this metastable minimum.
Ref.4 allows one spin flip, but never two or more at the same time, against
the energy, but keeps the free spins unchanged. This method works about
as good as the present one with respect to the final energy in two
dimensions; a combination of both tricks might work better but still
would not be exact. The genetic algorithm of ref.5 seems to circumvent even
these energy barriers for metastable states and may lead to the true
ground state.

Since submission of the paper, Gropengiesser [7] got better results by
combining the present flipping of free spins with the mutations of ref.4
to find the ground state energy, and on a parallel computer even got
superlinear speedup. A more mathematical and very efficient algorithm was
used by de Simone et al [8].

In summary, the fraction of never flipped spins only in one, five,
and six dimensions behaved qualitatively as in ferromagnets. But this
simple zero-temperature Glauber algorithm adapted from ref.3 seems to
be a way to get good though not exact ground state energies with moderate
computational effort even for large lattices.

We thank J.Kert\'esz for suggesting to check the fraction of never flipped
spins, and the Brazilian Education Ministry for a CAPES travel grant to DS.
\vfill \eject
$^*$ Present and permanent address: Institute for Theoretical Physics,
Cologne University, D-50923 K\"oln, Germany

[1] K.Binder and A.P. Young, Rev.Mod.Phys. 58, 801 (1986)

[2] S.A.Kauffman, {\it The Origins of Order}, Oxford University Press,
New York 1993

[3] E.S.Rodrigues and P.M.C. de Oliveira, J.Stat.Phys. 74, 1265 (1994)

[4] D.Stauffer, J.Stat.Phys. 74, 1293 (1994); M.Stevens, M.Cleary, and
D.Stauffer, Physica A 208, 1 (1994)

[5] P.Sutton, D.L.Hunter, and N.Jan, J.Physique I 4, 1281 (1994)

[6] B.Derrida, A.J.Bray, and C.Godreche, J.Phys. A 27, L 357 (1994);
D.Stauffer, J.Phys. A 27, 5029 (1994)

[7] U.Gropengiesser, J.Stat.Phys. 79, 1005 (1995) and Int.J.Mod.Phys. C,
in press

[8] C.de Simone, M.Diehl, M.J\"unger, P.Mutzel, G.Reinelt, G.Rinaldi,
J.Stat.Phys., in press.

\end